**Applying Predictive Machine Learning to Expand Flow in Container Yard Operations**

**Austin Ford Cooper**
Research Manager and Journal Manager
American Institute of Physics, Publishing
Melville, New York, 11747
Email: austinfcooper@gmail.com | acooper@aip.org

## Abstract

This study seeks to improve the throughput rates for shipping container terminals. In the United States, shipping ports link the domestic economy to global markets and are vital to sustain supply chain flow and economic stability. Maritime shipping accounts for nearly half of the U.S.'s annual international trade, two-thirds of which are represented by container shipping. Previous studies highlighted the capability of automation in enhancing container processing; however, unlike in European and East Asian ports, full automation is limited in U.S. ports due to legal protections for human labor. Consequently, there is a need for alternative methods that deliver automation-level efficiencies while maintaining the terms of cooperative agreements. This paper proposes an Intelligent Planning System (IPS) that applies the concept of Pareto Optimization to container yards through a mixed-integer linear programming (MILP) based recursive appointment system. The results show an improvement from baseline for both daily terminal throughput volumes and processing times. The generated IPS can be employed to provide recommendations for container positioning and truck pickup appointments to optimize container yard layout and flow resulting in reduced real-time congestion and predictively mitigated future congestion.

Congestion in port container yards is a complex, multi-variable problem that disrupts the supply chain. Container ports account for over $2.28 trillion, which equates to nearly half of the U.S.'s international trade (1). A 1% increase in port congestion in North America can raise international shipping freight cost-rates by approximately 0.5% (2). For an economic perspective, December 2021 shipping delays levels represented an increase of over 25% compared to 2016–2019 levels, estimated to be equivalent to a global ad-valorem tariff of 0.9–3.1% (3). The monetary impact of shipping rates at this December 2021 time is estimated to be $238 billion worth of cargo at two ports, Los Angeles and Long Beach, alone (4).

In recent years, research has focused on reducing congestion in container yards by exploring robotic and autonomous processes, larger gantry cranes, and physical expansion of port yards (5–7). However, many of these solutions are more applicable in regions like Europe and Asia, where labor practices permit greater use of autonomous systems (8). For example, China's plan for intelligent port upgrades aims to deploy over 6,000 autonomous container trucks by 2025, achieving approximately 20% automation in port logistics nationwide (9). Whereas in the USA, the International Longshoremen's Association's (ILA) Master Contract in the U.S. includes a provision that prohibits full automation of terminals, as of December 2024 (10).

Given the limitations on full automation in U.S. ports due to labor agreements, technologies such as planning systems play a crucial role in improving efficiency (11). However, majority of these technologies focus on vessel stowage planning, berth design and truck gate positioning (12). Exploration into the design and impact of predictive algorithms is heavily lacking in the research space. As a result, planning the optimal layout of container yards remains an understudied area, leaving significant potential for improvement in yard operations (13).

In this paper port terminals are treated as a unified ecosystem, where each component of the yard has individual peak efficiencies that when combined collectively optimize performance of the entire yard. The goal is to identify the ecosystem's collective, or Pareto, optimization level.

This study applies a Pareto optimization approach to container yard operations using a mixed-integer linear programming (MILP) and recursive booking system to recommend container positions and container-truck pickup appointments. By integrating z-score stacking principles with a collaborative appointment system, the proposed model aims to reduce processing and turnaround times through optimized yard layout. Moving from random to predictive, z-score-oriented container stacking offers a structured framework to enhance appointment efficiency, improving process times and overall terminal performance. The various elements of the programming are combined into a forward-looking, predictive software tool that enables yard planners to determine the optimal placement of both current and future containers. This combined framework is referred to as a predictive Intelligence Planning System (IPS).

Prior, relevant research studies and important definitions and concepts are discussed at length in the next section.

## Literature Review

Previous research has examined container yards as space-constrained areas where throughput is determined by the ability and reliability of drayage vehicles to enter and exit the yard efficiently for pickups (14-16).

Container yard layouts are designed to connect recently unloaded import containers, with drayage systems to transition containers to their final mode of transportation (17). Yards can be separated using various methods, such as separating empty and full containers, separating demurrage cargo from non-demurrage, or segmenting the yard by container content type (personal communication). Considerations for computationally driven solutions specific to the container terminal environment of the U.S. must include factors such as uneven technology adoption curves and non-universal yard layouts (18).

Despite variations in yard design across regions, the traditional procedure to container stacking follows that firstly, containers will occupy all available horizontal space after which they will be stacked vertically. The maximum stacking height is determined by each yard's terminal operator based on safety restrictions and the reach limits of equipment like reach stackers (17).

The most foundational stacking procedure is random container stacking. Yards that use random container arrangements position containers in the yard as they arrive, often without separating empty and full containers or without an ordering protocol. This approach works effectively in yards with ample space relative to the number of containers received. These yards can achieve higher throughput rates compared to space-constrained container yards (14).

[...] ner yards. **Figure A** portrays a [...] space constrained terminal.

**Container Terminal Comparison**

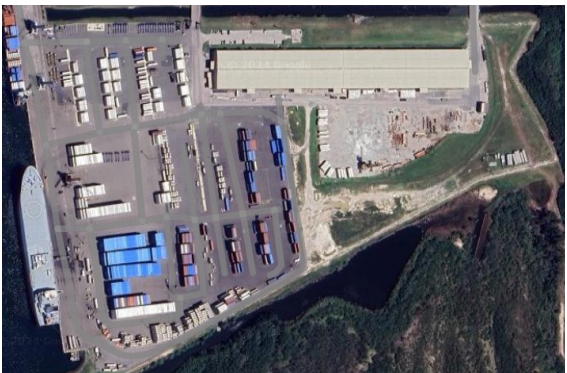
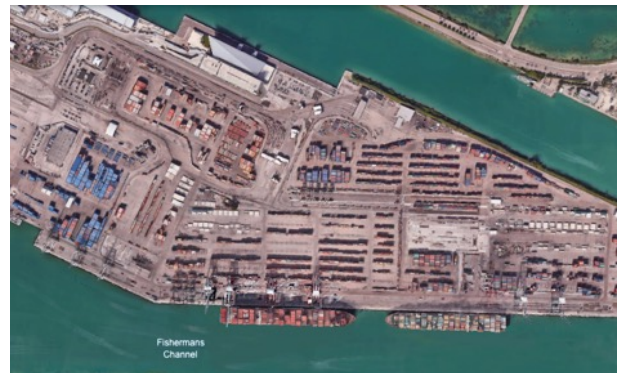

Figure A                                                     Figure B

**Figure I: Difference between space constrained and non-constrained container yards.**

Research into space-constrained container yards extends to non-randomized stacking processes also. The work of Kim et al. explored the application of truck appointment systems (TAS), generated by collaborative scheduling between trucking companies and container terminals to reduce peak-hour congestion and minimize the number of containers in the yard without pickup appointments (18). TAS usage in container terminals is increasing, but their adoption varies globally. For instance, all container terminals at the Los Angeles and Long Beach port complex have implemented TAS to manage truck flows, reduce congestion, and decrease waiting times. In contrast, adoption remains slower in other major port complexes, such as the Port Authority of New York and New Jersey, where only a few terminals have implemented these systems (19).

In their work, Hiroshi and Azab expanded the use of appointment scheduling systems to account for the vertical positioning of containers in the yard, highlighting the significant impact of multiple container moves on throughput times. Multiple container moves occur when the desired container is buried in a stack, requiring other containers to be relocated to access it (20). These additional movements not only increase yard processing times but also elevate the risks of damages, injuries, and associated financial costs. To address these challenges, Hiroshi and Azab modeled the container yard as a bi-objective integer optimization problem, incorporating container positions as a key factor to minimize disruptions and improve overall efficiency. Though their work raised a new tactic to consider multiple quantitative factors, the behavioral elements of the supply chain were unaddressed.

In addition to quantitative factors as explored by Hiroshi and Azab, container terminals also face qualitative challenges that must be factored into optimizations. One of the latest advancements for incorporating qualitative variables into system performance modeling is the work of Romero et al. Their research addressed the complex layout of wind farms by accounting for behavioral dynamics and performance metrics (21). They developed a model that translated a bi-objective integer framework into a mixed-integer linear and non-linear programming model. The model, however, is limited in scope, only applied wind farms and lacking exploration of the impact in other industries such as shipping.

The strategic selection of a container's initial vertical position is essential to developing a truly accurate container yard model. Ting et al. analyzed vertical stacking positions to optimize onset container placement and reduce the number of multiple moves required for pickups (15). Their approach introduced the principle of Z-stacking, which determines a container's position in a vertical stack based on factors such as container contents, arrival date, and pickup appointment date. The full list of factors used to determine the Z position in the stack are detailed in **Table I**.

|  |  | **Z-score factors** |
|---|---|---|
| Date of issuing arrival notice | Gross container weight | Consignee/carrier |
| Container contents | Existing truck pickup appointment status | Free days remaining / Demurrage status |
| Delivery date/time | Shipper | Final destination |

**Table I: Z-score factors**

Under Z-stacking, stacking positions are determined using discriminant analysis. Linear discriminant functions are derived to represent the boundaries between a category for consignee and cargo types and the remaining two categories. These functions are shown in **Equations 1, 2 and 3**. The discriminant Z-scores for categories C1, C2, C3 are calculated as weighted measures, forming the basis for classifying containers into one of the three categories (15).

$$C1 = -0.985 + 0.032 \text{ x Consignee} + 1.281 \text{ x Cargo} \qquad (1)$$
$$C2 = -13.239 + 0.116 \text{ x Consignee} + 4.698 \text{ x Cargo} \qquad (2)$$
$$C3 = -37.387 + 0.344 \text{ x Consignee} + 7.688 \text{ x Cargo} \qquad (3)$$

This paper builds on the prior research in maritime dynamics and computational optimization by treating container yards as continuous systems where the flow of containers in and out can be optimized. The following section outlines the methodologies used to evaluate the impact of predictive intelligence planning on throughput and processing time.

## Methods

This work's approach introduces category stacking through Z-scores, integrated with TAS supported MILP analysis. This solution identifies the optimal operational rate for the container yard and generates placement recommendations to maintain this rate consistently.

The IPS study begins by addressing decongestion through upward scaling of pickup appointments to optimize turnaround times and eliminate "traffic jams."

Next, we streamline the planning process for container placement. Using the MILP algorithm's output to account for new container arrivals, and current and future appointments, to ensure the fastest turnaround times.

To further enhance yard efficiency, we reduce the number of containers without pickup appointments by implementing a TAS designed to sustain the maximum serviceable flow rate of the container yard throughout all hours of operation.

The following variables were used to calculate the available space for container yard segmentation:

|  |  | Factors for yard segmentation |
| --- | --- | --- |
| Yard length | Yard width | Total number of containers in yard |
| Entry gate location | Exit gate location | |

**Table II: Factors for yard segmentation**

## Equations for MILP Modeling
### Z-Score Calculations

Container positions under MILP begin with determining discriminate Z-scores using equations 1, 2 and 3. All three equations use variables for Cargo and Consignee. These variables are discussed in the below.

### Cargo Variable

The *Cargo* variable is based on the contents of the container. The cargo variable for each container is calculated by **Equation 4.**

$$C_v = w_c \times p_c \times d_f \qquad (4)$$

Where:
$C_v$ = Cargo variable
$w_c$ = Container weight
$p_c$ = Cargo percentage
$d_f$ = Number of remaining freedays for the container

Cargo percentages reflect the likelihood that the container will be picked up within a given time frame, as determined in the work of Ting et al. (17).

*Consignee Variable*

In MILP the *Consignee* variable is a continuous calculation. The *Consignee* variable accounts for prior interactions and is calculated through the following steps (17):

1) Days Passed Since Arrival ($d_p$):
   The difference between the current date ($d_{current}$) and the container's arrival date ($d_{arrival}$), as given in **Equation 5**.

$$d_p = d_{current} - d_{arrival} \qquad (5)$$

   Where:
   $d_p$ = Container days passed
   $d_{current}$ = Current date
   $d_{arrival}$ = Arrival date

2) Remaining Free Days ($REM_f$):
   Calculated by subtracting ($d_p$) from the initial free days ($d_f$), as shown in **Equation 6**.

$$REM_f = d_f - d_p \qquad (6)$$

   Where:
   $REM_f$ = Remaining free days
   $d_f$ = Inital free day count
   $d_p$ = Container days passed

3) Consignee Variable ($C_{consignee}$):
   The final consignee variable is derived as shown in **Equation 7.** Within the calculation, the number of visits a trucking company makes per month is captured through the truck appointment system.

$$C_{consignee} = \frac{\left(\frac{1}{d_f}\right) \times REM_f}{S_v} \qquad (7)$$

   Where:
   $d_f$ = Inital free day count
   $REM_f$ = Remaining free days
   $S_v$ = trucking company visits per month

If a container does not have an appointment for a truck pickup instead of focusing on the trucking company, the calculation considers the total number of containers belonging to the same owner (i.e. shipping company) that are scheduled to arrive at the terminal without an appointment during the same month.

*Operational Process Times*

Average equipment times, gate clearance durations, and other processing variables were
collected through interviews with port terminal operators.

To evaluate the outputs of the MILP and identify potential improvements, the relative processing
times of a hypothetical container yard were analyzed. These performance measures and
associated calculations are governed by the internal operational process time (IO) and departure
time (DT).

1) Departure Time (DT):
   The departure time is determined using **Equation 8**.

$$DT = \frac{M_0 \text{ or } M_t}{S_{\text{gate}}} \times T_{\text{clear}} \tag{8}$$

   Where:
   $DT$ = Departure Time
   $M_0$ = Maximum truck count able to wait in the yard
   $M_t$ = Current truck count in the yard
   $S_{\text{gate}}$ = Maximum number of trucks serviceable simultaneously at clearance gate
   $T_{\text{clear}}$ = Time for a single truck to clear the departure gate

2) Internal Operational Process Time (DT):
   The internal operational process time is determined using **Equation 9**.

$$IO = T_{\text{load}} + T_{\text{inspect}} \tag{9}$$

   Where:
   $IO$ = Internal Operational Process Time
   $T_{\text{load}}$ = Time for truck to be loaded with a container
   $T_{\text{inspect}}$ = Time for safety inspections

3) **Processing Time** (PT):
   Total processing time combines DT and IO, as shown in **Equation 10**.

$$PT = DT + IO \tag{10}$$

Where:
DT = Processing Time
DT = Departure Time
IO = Internal Operational Process Time

As trucks enter the yard during the calculated time interval block, the following process is
applied:

1. The departure time (DT) for the current time block is calculated, replacing $M_0$ with $M_t$ .

2. If DT>IO, it indicates that congestion is occurring or will occur.
   a. Under the appointment system, trucks are shifted to a new time block or appointment window to alleviate congestion.

By structuring appointments so that $M_t$ is never significantly above or below $M_0$, it is possible to recover truck counts lost during previously underutilized time blocks, and simultaneously reduce congestion in overburdened time blocks.

*Throughput and Optimization Metrics*

1) Total Throughput ($T_{\text{throughput}}$):
   Throughput improvements are calculated using **Equation 11**.

$$T_{\text{throughput}} = \frac{M_{\text{optimized}} - M_{\text{baseline}}}{M_{\text{baseline}}} \tag{11}$$

Where:
   $T_{\text{throughput}} =$ Total throughput
   $M_{\text{optimized}} =$ Total trucks serviced in optimized real time
   $M_{\text{baseline}} =$ Total trucks serviced in baseline conditions

2) Processing Time Improvements ($PT_{\text{improve}}$):
   Processing time improvements are determined using **Equation 12**.

$$PT_{\text{improve}} = \frac{\left|PT_{\text{hyp}} - PT_{\text{real}}\right| - \left(PT_{\text{hyp}} - PT_{\text{baseline}}\right)}{PT_{\text{hyp}}} \tag{12}$$

Where:
   $PT_{\text{improve}} =$ Processing time improvement
   $PT_{\text{hyp}} =$ Hypothetical optimized processing time
   $PT_{\text{real}} =$ Real-time optimized processing time
   $PT_{\text{baseline}} =$ Baseline processing time

Positive $PT_{\text{improve}}$ indicates a decrease in M count is needed to decrease congestion. Negative $PT_{\text{improve}}$ indicates the yard is not operating at peak capability and in conjunction with increasing M counts there is room for the processing time to be made greater before returns diminish.

*Implementation*

The MILP algorithm and predictive software were developed using Python and connected to a front-end web application hosted on a local cloud drive. Visualizations utilized FreeCAD. Calculations were performed on a system with a 2.9 GHz dual-core Intel Core i5 processor and 8 GB of 2133 MHz LPDDR3 memory.

Utilizing the process outlined in the flow diagram of **Figure II** the MILP algorithm and recursive appointment system were run on an experimental data set uploaded through a CSV file, representative of the output to be retrieved from a port terminal operating system (TOS) through an API call.

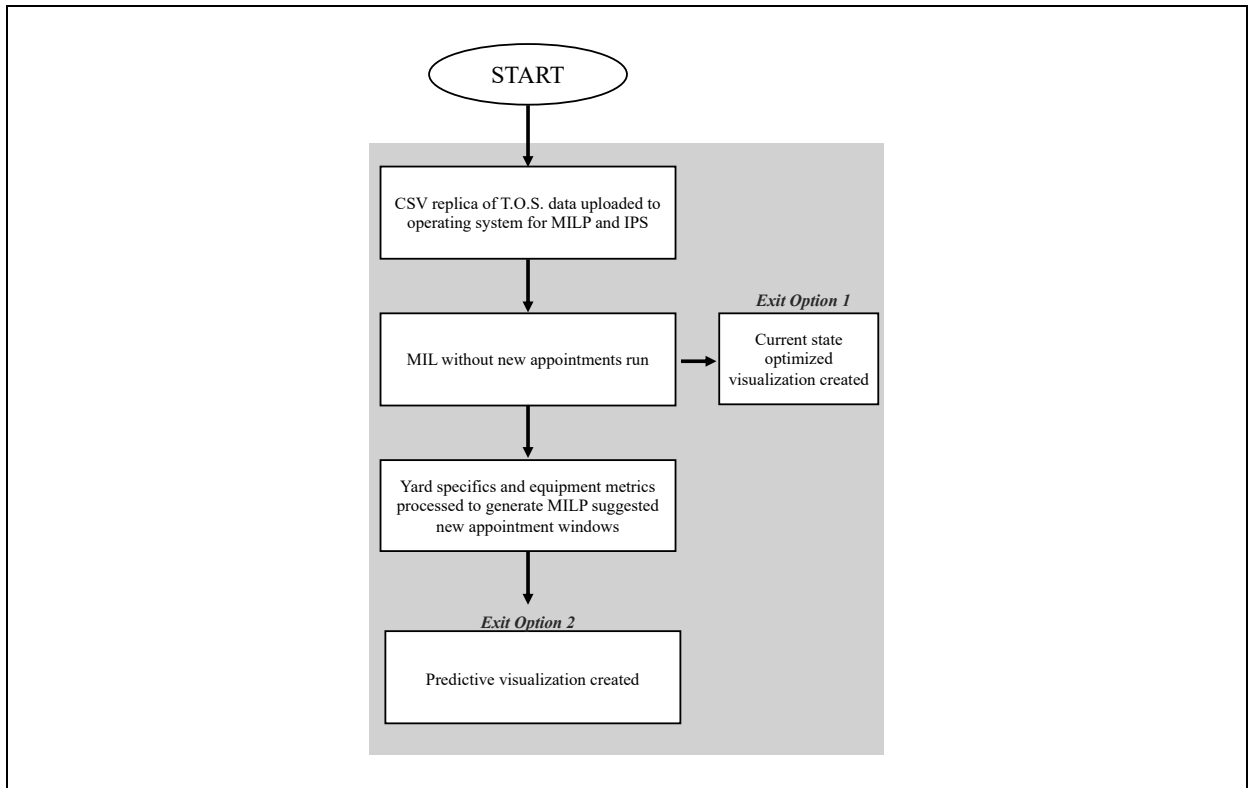

**Figure II: Process for testing the MILP algorithm and recursive appointment system.**

## Results and Discussion

The MILP algorithm and visualization software were applied to an experimental dataset. The experimental dataset included 63 sample containers with complete data fields. Four outputs were generated using the visualization software to demonstrate the combined impact of the Z-score stacking through MILP and recursive TAS appointments for yard optimization.

Insights from interviews with port operators revealed three primary categorizations for containers:

1. Containers within the free day period with scheduled pickup appointments.
2. Containers within the free day period without scheduled pickup appointments.
3. Containers outside the free day period without scheduled pickup appointments.

Containers are assigned to one of the three above categories. Within each category, the Z-score stacking principle is applied. When new appointments are created, the MILP algorithm determines the optimal placement for the container within the existing optimized yard layout. All outputs are based on the same experimental CSV dataset, and the yard layouts and their respective outputs are presented in **Table III**.

| | Yard Segmentations Analyzed |
|---|---|
| Output 1 | Randomized Stacking, No Segmenting |
| Output 2 | Randomized Stacking, Segmenting |
| Output 3 | Z-Score Stacking, Segmenting |
| Output 4 | (IPS) - Z-Score Stacking, Segmenting, Recursive Appointment Scheduling |

**Table III: Yard Outputs**

For all four outputs, the yards' Relative Operating Time and Total Daily Container Throughputs (M) were calculated and are shown in **Figure III** below.

| | | Outputs from IPS |
|---|---|---|
| 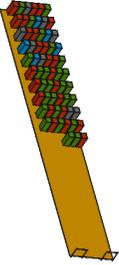 Yard I isometric left view | 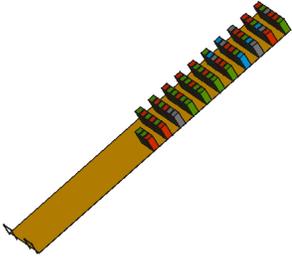 Yard I isometric right view | **Yard I: Container yard that maximizes yard space, Randomized stacking (stacking not used)**<br><br>Values treated as the baseline levels for yard comparison.<br><br>**Processing Time (PT):** 370<br>**Throughput (M):** 37 |
| 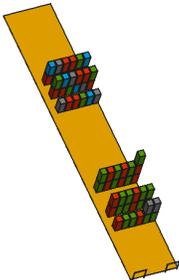 Yard II isometric left view | 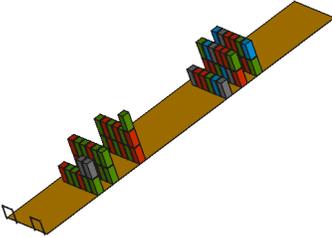 Yard II isometric right view | **Yard II: Container Yard with Segmentation and randomized stacking**<br><br>Segmentation allows for the baseline level throughput, M, to be achieved in expedited processing time.<br><br>**Processing Time (PT):** 330<br>**Throughput (M):** 37 |
| 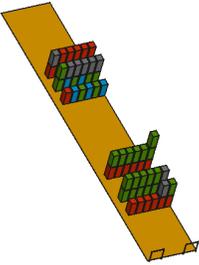 Yard III isometric left view | 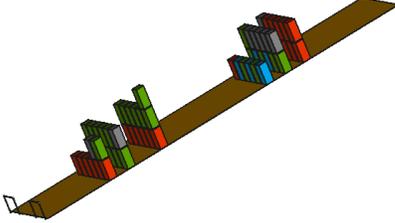 Yard III isometric right view | **Yard III: Container Yard with Segmentation and Z-Score Stacking**<br><br>The greatest processing time and container throughput possible without adding new appointments or reallocating higher than congestion-threshold trucks.<br><br>**Processing Time (PT):** 330<br>**Throughput (M):** 49 |

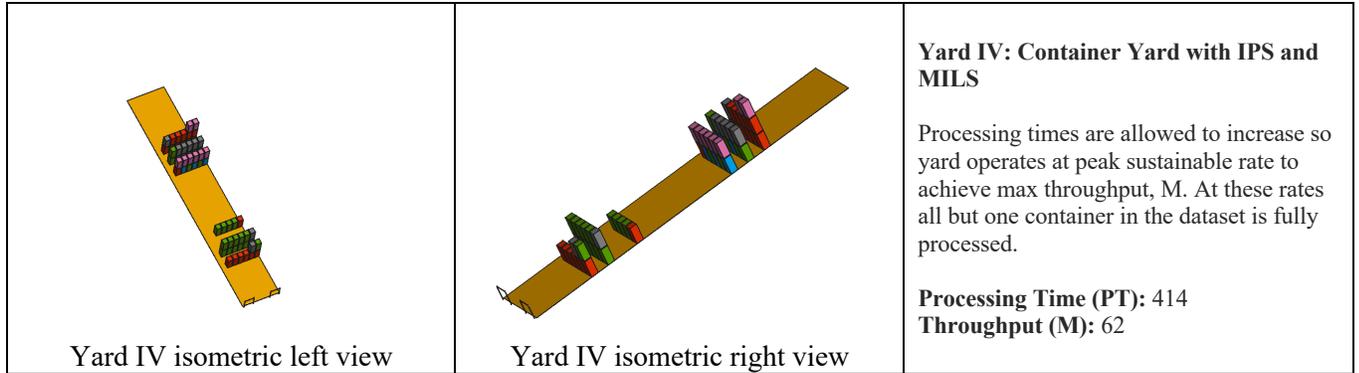

| Yard IV isometric left view | Yard IV isometric right view | **Yard IV: Container Yard with IPS and MILS**<br><br>Processing times are allowed to increase so yard operates at peak sustainable rate to achieve max throughput, M. At these rates all but one container in the dataset is fully processed.<br><br>**Processing Time (PT):** 414<br>**Throughput (M):** 62 |

**Figure III: Relative Operating Time and Daily Container Throughputs achieved under each output**

To assess the maximum impact achievable through the IPS framework, the results of Yard 1 and Yard 4 are discussed in detail. Yard 1 exemplifies a congestion-prone scenario, characterized by randomized stacking and minimal use of appointments, while Yard 4 reflects the ideal implementation of all proposed strategies, including categorization, z-score stacking, and optimized appointment systems.

**Yard I** represents a standard container yard that employs randomized stacking without any segmentation. Containers are arranged based solely on available yard space, as such in this instance vertical stacking is not used, yet containers are still horizontally positioned in front of others causing impedance. The lack of structured stacking or segmentation results in frequent, multiple container movements to access desired containers; increasing processing times, exacerbating congestion, and reducing the maximum achievable container flow rates.

**Yard IV**, in contrast to **Yard I,** incorporates categorization, segmentation, and Z-score-based stacking. This approach reduces processing time per container by enabling shorter, more direct retrieval and loading operations while increasing the number of simultaneous appointment windows that can be offered. Using Z-score stacking based on scheduled pickup dates, minimizes multiple container moves and enhances maximum container flow rates. Yard IV positions containers using the MILP algorithm and recursive TAS appointments. The MILP algorithm ensures that container repositions are minimized, preserving the efficiency of existing placements.

In **Yard IV**, containers are segmented as follows:
1. **Section 1**: Pre-existing pickup appointments (non-demurrage containers).
2. **Section 2**: Non-demurrage containers without appointments.
3. **Section 3**: Demurrage containers without appointments and newly scheduled appointments for demurrage containers (differentiated with pink color coding).

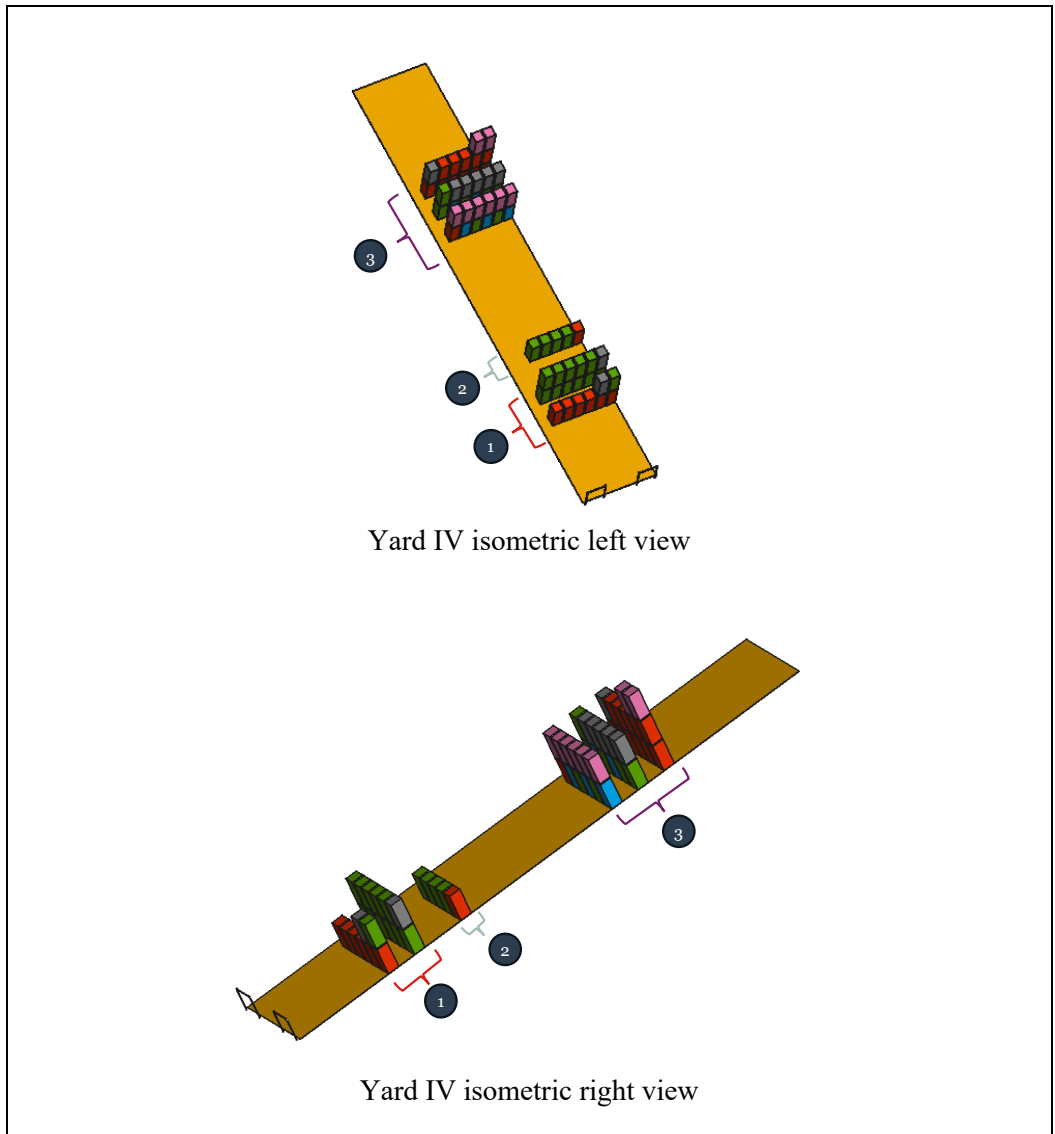

Yard IV isometric left view

Yard IV isometric right view

**Figure III:** illustrates the segmentation used in **Yard IV**
.

Every container yard has an upper limit on the number of trucks it can process per hour. Trucks exceeding this threshold cause congestion and reduce daily throughput. Terminals operating without appointments, will see select hours with truck volumes exceeding the threshold, while other hours fall far below the threshold. This results in unoptimized container movement and lower-than-possible daily throughput.

    **Figure IV** applies Departure Time (DT) and Internal Operational Process Time (IO) measurements (as defined by **Equations 11 and 12**) to the hypothetical container yard. Stopping conditions for the yard were set using **Equations 8 and 9**, with an M-count threshold of 60, reflecting typical daily counts at a terminal processing 160,000 TEUs annually. The blue bars depict the current M-count attempting to access the yard during each hour of a 9-hour operational day, while the red line indicates the stopping threshold where excess M-counts contribute to congestion.

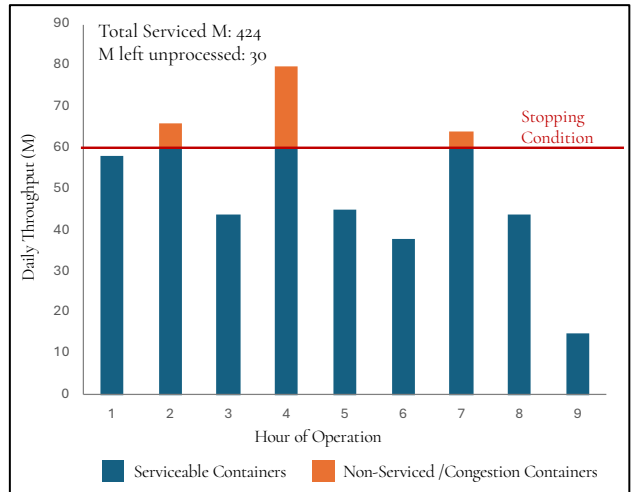

**Figure IV: DT and IO Equations Applied to a Container Yard Without MILP/IPS.**

By employing an IPS, congestion-causing trucks can be shifted from overburdened time windows to underutilized ones. Additional trucks can also be added to fill any remaining underutilized windows. This improvement is illustrated in **Figure V.**

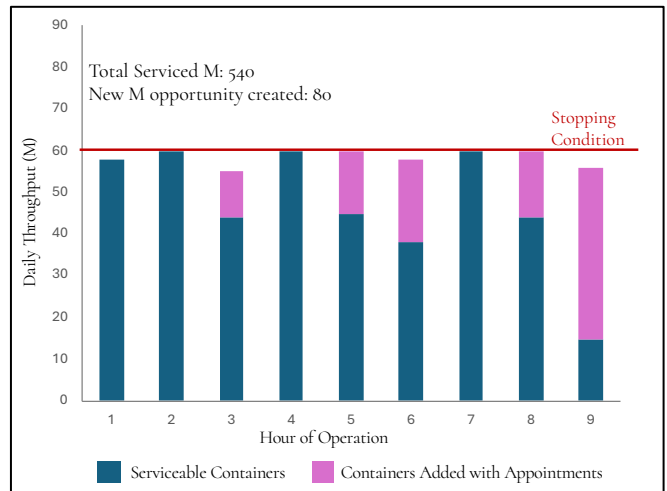

**Figure V: DT and IO Equations Applied to a Container Yard With MILP/IPS.**

The implementation of an IPS leads to improved truck turnaround times for existing pickup appointments, a reduced number of containers without appointments, streamlined planning processes with yard placement recommendations from the algorithm, and better-positioned containers for future pickups and arrivals.

Results from this study suggest implementing an IPS can improve processing times and total daily movable container counts by 10–20%. For a yard with the capacity to process 160,000 TEUs annually, this translates to accommodating one additional vessel every five weeks. Based on industry pricing averages, this improvement corresponds to between $3.74 million - $7 million in additional monthly revenue (22–25).

These findings support the hypothesis that transitioning container yards from random stacking to predictive, Z-score-oriented categorization, combined with a recursive appointment booking system, significantly reduces processing and turnaround times, thereby maximizing throughput.

## Future Considerations

The discussed MILP and Z-score categorization-based container assortment system, combined with a recursive TAS, demonstrates the potential to significantly enhance container yard operations. By reducing congestion and increasing daily container flow, the system can not only improve processing times, but it also enables terminal operators to generate additional revenue through the accommodation of new vessel berths. Further advancements can be achieved with deeper algorithmic integration, such as leveraging convolutional neural networks (CNNs) trained on real-time yard metrics and activities for greater predictive accuracy in container positioning.

The study introduced expands upon prior investigations into port terminal dynamics, treating the yard as an interconnected ecosystem where operational components, equipment, and behavioral dynamics are viewed as non-linear variables impacting overall system performance. By drawing direct parallels to Pareto Optimization principles used in manufacturing, the MILP framework offers a structured, forward-looking planning tool that could be integrated into existing software platforms. This would allow yard planners to make intelligent, data-driven container arrangements that optimize current operations while preparing for future demands.

The proposed system introduces a mathematical model capable of continuous operation when connected to a terminal operating system (TOS). The introduced software operates on uploaded CSV data, providing an effective method for generating container yard optimizations. However, for a machine learning algorithm to continuously generate optimizations and predictions, the ideal configuration involves API integration with the terminal operating system (TOS). While CSVs or similar data uploads can serve as functional inputs, they offer only snapshot optimizations rather than continuous, real-time predictions. By integrating real-time data, this model can evolve into an artificial neural network, forming a dynamic intelligent planning system. This IPS provides terminal operators with efficiency levels comparable to those achieved by autonomous robotic systems, while preserving the existing workforce's role. Moreover, the IPS framework is adaptable, ensuring a smooth transition to autonomy, should a terminal choose to adopt additional automated processes.

Building on this adaptable framework, the proposed IPS not only enhances terminal efficiency but also ensures compatibility with future advancements in automation.

Two critical considerations warrant further near-term exploration:

**Economic feasibility**: Whereas certain studies have explored the economic impact of appointment systems on driver dispatching or the tradeoffs of assigning penalties to carriers for missed pickup appointments, the introduced IPS focuses on enhancing revenue generation by enabling terminals to berth more vessels annually. A promising direction for future research involves integrating real-time data from terminal operating systems (TOS) to quantify the tangible increase in additional TEUs processed daily. Furthermore, the introduction of an IPS-centric role for clerks and dockworkers may create new lower physically intensive career pathways, aligning with workforce development goals and advancing the operational efficiency of port terminals.

**Complete life cycle assessment (LCA)**: It is now imperative to quantify the full LCA of an IPS; encompassing the benefits of optimized container flows, enhanced job roles, and increased terminal throughput. Revenue generation potential, along with frameworks for further technological integration, should be evaluated in tandem with the organizational change processes required for successful implementation. The assessment should include not only direct impacts but also long-term scalability and adaptability within diverse operational contexts.

By addressing these considerations, the IPS represents a significant step forward in achieving sustainable, data-driven container yard optimization. Future advancements in this field should continue to bridge the gap between theoretical models and practical applications, ensuring that ports remain competitive and resilient in the face of evolving global supply chain demands.

## Conclusion

This study demonstrates the potential of combining machine learning algorithms and MILP computations to address congestion and enhance container throughput in U.S. port terminals. This research highlights the importance of a data-centric framework that respects the cooperative agreements and economic motivations unique to U.S. ports.

By integrating engineering principles, machine learning techniques, and practical industry insights, this approach offers a robust solution to maintain competitiveness and improve operations. The proposed IPS and MILP computational methods represent a holistic, interdisciplinary solution that amplifies terminal capabilities while aligning with the long-term vision of sustainable and efficient port operations.

## Author Contributions

The authors confirm contribution to the paper as follows: study conception and design: A.F. Cooper; data collection: A. F. Cooper; analysis and interpretation of results: A.F. Cooper; draft manuscript preparation: A.F. Cooper. All authors reviewed the results and approved the final version of the manuscript.